\documentclass[twocolumn, superscriptaddress, amsmath,amssymb, prc, aps]{revtex4-1}

\usepackage{graphicx}
\usepackage{times}
\usepackage{xcolor}
\usepackage{bm}
\usepackage{hyperref}

\usepackage{url}

\begin{document} 

\title{New Constraints on Axion-Like Particles with the NEON Detector at a Nuclear Reactor}

\author{Byung Ju Park}
\affiliation{IBS School, University of Science and Technology (UST), Deajeon 34113, Republic of Korea}
\affiliation{Center for Underground Physics, Institute for Basic Science (IBS), Daejeon 34126, Republic of Korea}
\author{Jae Jin Choi}
\affiliation{Department of Physics and Astronomy, Seoul National University, Seoul 08826, Republic of Korea}
\affiliation{Center for Underground Physics, Institute for Basic Science (IBS), Daejeon 34126, Republic of Korea}
\author{Eunju Jeon}
\affiliation{Center for Underground Physics, Institute for Basic Science (IBS), Daejeon 34126, Republic of Korea}
\affiliation{IBS School, University of Science and Technology (UST), Deajeon 34113, Republic of Korea}
\author{Jinyu Kim}
\affiliation{HANARO Utilization Division, Korea Atomic Energy Research Institute (KAERI), Deajeon 34057, Republic of Korea}
\author{Kyungwon Kim}
\affiliation{Center for Underground Physics, Institute for Basic Science (IBS), Daejeon 34126, Republic of Korea}
\author{Sung Hyun Kim}
\affiliation{Center for Underground Physics, Institute for Basic Science (IBS), Daejeon 34126, Republic of Korea}
\author{Sun Kee Kim}
\affiliation{Department of Physics and Astronomy, Seoul National University, Seoul 08826, Republic of Korea}
\author{Yeongduk Kim}
\affiliation{Center for Underground Physics, Institute for Basic Science (IBS), Daejeon 34126, Republic of Korea}
\affiliation{IBS School, University of Science and Technology (UST), Deajeon 34113, Republic of Korea}
\author{Young Ju Ko}
\affiliation{Center for Underground Physics, Institute for Basic Science (IBS), Daejeon 34126, Republic of Korea}
\author{Byoung-Cheol Koh}
\affiliation{Department of Physics, Chung-Ang University, Seoul 06973, Republic of Korea}
\author{Chang Hyon Ha}
\affiliation{Department of Physics, Chung-Ang University, Seoul 06973, Republic of Korea}
\author{Seo Hyun Lee}
\affiliation{IBS School, University of Science and Technology (UST), Deajeon 34113, Republic of Korea}
\affiliation{Center for Underground Physics, Institute for Basic Science (IBS), Daejeon 34126, Republic of Korea}
\author{In Soo Lee}
\email{islee@ibs.re.kr}
\affiliation{Center for Underground Physics, Institute for Basic Science (IBS), Daejeon 34126, Republic of Korea}
\author{Hyunseok Lee}
\affiliation{IBS School, University of Science and Technology (UST), Deajeon 34113, Republic of Korea}
\affiliation{Center for Underground Physics, Institute for Basic Science (IBS), Daejeon 34126, Republic of Korea}
\author{Hyun Su Lee}
\email{hyunsulee@ibs.re.kr}
\affiliation{Center for Underground Physics, Institute for Basic Science (IBS), Daejeon 34126, Republic of Korea}
\affiliation{IBS School, University of Science and Technology (UST), Deajeon 34113, Republic of Korea}
\author{Jaison Lee}
\affiliation{Center for Underground Physics, Institute for Basic Science (IBS), Daejeon 34126, Republic of Korea}
\author{Yoomin Oh}
\affiliation{Center for Underground Physics, Institute for Basic Science (IBS), Daejeon 34126, Republic of Korea}
\collaboration{NEON Collaboration}
\author{Doojin Kim}
\affiliation{Department of Physics, University of South Dakota, Vermillion, SD 57069, USA}
\affiliation{Mitchell Institute for Fundamental Physics and Astronomy, Department of Physics and Astronomy,Texas A\&M University, College Station, TX 77845, USA} 
\author{Gordan Krnjaic}
\affiliation{Theoretical Physics Division, Fermi National Accelerator Laboratory, Batavia, IL, USA}
\affiliation{Department of Astronomy \& Astrophysics, University of Chicago, Chicago, IL USA}
\affiliation{Kavli Institute for Cosmological Physics, University of Chicago, Chicago, IL USA}
\author{Jacopo Nava}
\affiliation{Dipartimento di Fisica e Astronomia, Università di Bologna, via Irnerio 46, 40126, Bologna, Italy}
\affiliation{INFN, Sezione di Bologna, viale Berti Pichat 6/2, 40127, Bologna, Italy}
\date{\today}

\begin{abstract} 
We report new constraints on axion-like particles (ALPs) using data from the NEON experiment, which features 16.7\,kg of NaI(Tl) target located 23.7\,m from a 2.8\,GW thermal power nuclear reactor. Analyzing a total exposure of 3063\,kg$\cdot$days, with 1596\,kg$\cdot$days during reactor-on and 1467\,kg$\cdot$days during reactor-off periods, we compared energy spectra to search for ALP-induced signals.    
No significant signal was observed, enabling us to set exclusion limits at the 95\% confidence level. These limits probe previously unexplored regions of the ALP parameter space, particularly for axion masses (m$_a$) near 1\,MeV/c$^2$. For ALP-photon coupling (${g_{a\gamma}}$), limits reach as low as 6.24$\times$ 10$^{-6}$\,GeV$^{-1}$ at m$_a$ = 3.0\,MeV/c$^2$, while for ALP-electron coupling (${g_{ae}}$), limits reach  4.95$\times$ 10$^{-8}$ at m$_a$ = 1.02\,MeV/c$^2$. 
This work demonstrates the potential for future reactor experiments to probe unexplored ALP parameter space. 
\end{abstract}

\maketitle

Axions were first proposed in 1977 by Peccei and Quinn~\cite{Peccei:1977hh} to address the strong \textit{CP} problem in quantum chromodynamics (QCD)~\cite{Wilczek:1977pj,Weinberg:1977ma}. Due to their extremely light mass and weak interactions with ordinary matter, axions are considered promising candidates for dark matter~\cite{Kim:1979if,Shifman:1979if,Preskill:1982cy,Abbott:1982af,Dine:1982ah}. Despite numerous experimental searches, axions have not yet been detected~\cite{ADMX:2021nhd,HAYSTAC:2020kwv,CAPP:2020utb,ParticleDataGroup:2022pth}. 
The concept has since been extended to include axion-like particles (ALPs) in various models~\cite{Witten:1984dg,Choi:2020rgn}. While ALPs share many properties with axions, making them viable dark matter candidates, they are not necessarily tied to solving the strong \textit{CP} problem~\cite{Jaeckel:2010ni}. ALPs can span a wide range of masses and coupling constants, leading to diverse phenomenological implications in astrophysical and laboratory contexts~\cite{Choi:2020rgn}.  

ALPs interact with Standard Model particles, particularly photons and electrons, motivating extensive experimental searches~\cite{ParticleDataGroup:2022pth}. Light ALPs (axion mass, m$_{a}<$~100\,keV/c$^2$) are typically investigated using solar helioscopes, haloscopes, or photon regeneration experiments~\cite{Semertzidis:2021rxs}. In contrast, heavy ALPs (m$_{a}>$~100\,keV/c$^2$) are probed using colliders or beam dump experiments~\cite{Jaeckel:2015jla,Andreev:2021fzd,Capozzi:2023ffu}. Astrophysical observations provide complementary constraints on the ALP parameter space~\cite{Raffelt:2006cw,Caputo:2024oqc}. 

A particularly intriguing  region of ALP parameter space, known as  the  ``cosmological triangle''~\cite{Brdar:2020dpr,Lucente:2022wai}, spans masses between 0.3 and 8\,MeV/c$^2$ with an axion-photon coupling constant ($g_{a\gamma}$) ranging from  1.8$\times$10$^{-6}$ to 5$\times$10$^{-5}$\,GeV$^{-1}$. This region remains largely unexplored by both direct searches and astrophysical bounds. 
Although model-dependent cosmological constraints~\cite{Depta:2020zbh} restrict this region, they can be evaded under nonstandard cosmological scenarios~\cite{Carenza:2020zil,Depta:2020wmr}.

Recent supernova-based constraints from Ref.~\cite{Caputo:2021rux} suggest coverage of this region; however, these results depend on specific assumptions about the muonic supernova core and the Garching supernova model~\cite{Janka:2017vcp,Bollig:2020xdr}. 
Notably, ALP production and the corresponding constraints depend strongly on supernova mechanisms and ALP model parameters. Recent studies suggest that in certain regimes, loop-induced ALP-photon interactions may dominate over muonic processes, potentially altering the derived limits~\cite{Ferreira:2022xlw}. 

Growing interest in this region~\cite{Lucente:2022wai} has driven studies exploring the potential for direct ALP searches with significantly less model dependence. These include short-baseline reactor experiments~\cite{PhysRevLett.124.211804,AristizabalSierra:2020rom}, the accelerator-based Coherent CAPTAIN-Mills~(CCM) experiment with a 10-ton liquid argon target~\cite{CCM:2021jmk}, DUNE-like future neutrino experiments with a 50-ton liquid or gaseous argon target~\cite{Brdar:2020dpr}, and a 2-kton liquid scintillator with an intense proton beam underground~\cite{Seo:2023xku}.

Nuclear reactors are the most intense sources of photons with energies up to a few MeV. Since ALPs can be produced via photon-induced scattering~\cite{Choi:2020rgn}, reactors offer a promising avenue for ALP searches in the MeV/c$^2$ mass range. However, only a few reactor-based ALP searches have been conducted~\cite{TEXONO:2006spf,Sahoo:2024zee}. 
In reactor-based ALP searches, data collected during reactor operation (reactor-on data) can be compared to data collected when the reactor is inactive (reactor-off data) to constrain potential ALP signals. 
ALPs are primarily produced via Primakoff or Compton-like processes and detected through decay or scattering channels. 

In this study, we present a direct search for ALPs using the NEON (Neutrino Elastic scattering Observation with NaI) experiment~\cite{NEON:2022hbk}. 
Leveraging the intense ALP production from the reactor core, the NEON experiment begins to probe the  unexplored ``cosmological triangle'' in a laboratory-based experiment. For axion-electron couplings, this study investigates previously uncharted parameter space for axion masses between 300\,keV/c$^2$ and 1\,MeV/c$^2$.

The NEON  experiment is designed to detect coherent elastic neutrino-nucleus scattering (CE$\nu$NS) using reactor electron antineutrinos~\cite{NEON:2022hbk}. The detector is located in the tendon gallery of the 2.8\,GW Hanbit nuclear power reactor, 23.7 $\pm$ 0.3\,m from the center of the reactor core. 
After an engineering run in 2021~\cite{NEON:2022hbk}, the detector encapsulation was upgraded to enhance long-term operational stability~\cite{NEON:2024Upgrade}.

The NEON detector consists of four 8-in. and two 4-in. long, 3-in. diameter NaI(Tl) crystals, with a total mass of 16.7\,kg. 
The six NaI(Tl) modules are submerged in 800\,l of liquid scintillator. This liquid scintillator helps identify and reduce radioactive backgrounds affecting the NaI(Tl) crystals~\cite{Adhikari:2020asl}. To further reduce external radiation background, the liquid scintillator is surrounded by shielding made of lead, borated polyethylene, and high-density polyethylene~\cite{NEON:2022hbk}.

Each NaI(Tl) crystal is coupled to two photomultiplier tubes (PMTs) without quartz windows, optimizing  light collection efficiency~\cite{Choi:2020qcj,NEON:2024Upgrade}. The crystal-PMT assemblies are enclosed in a copper casing to ensure structural integrity and prevent exposure to external air or liquid scintillator~\cite{NEON:2024Upgrade}. Events that satisfy the trigger condition--coincident photoelectrons detected by both of the crystal's PMTs within a 200\,ns window--are recorded using 500\,MHz flash analog-to-digital converters (FADCs). These events are stored as 8\,$\mu$s waveforms, beginning 2.4\,$\mu$s before the trigger~\cite{Adhikari:2018fpo,NEON:2022hbk}. 
The system records two readouts: a high-gain signal from the anode for the 0--60\,keV energy range and a low-gain signal from the fifth-stage dynode for the 60--3000\,keV range, similar to the setup used in the COSINE-100 experiment~\cite{Adhikari:2018fpo}.
To reject unwanted phosphorescence events from direct muon hits, a 300\,ms dead time is applied for energy deposits exceeding approximately 3\,MeV.

The data analyzed in this study were collected between April 11, 2022 and June 22, 2023, yielding a total live time exposure of 5702 kg$\cdot$days.
Data collection was generally stable, although downtime occurred due to unexpected power outages. 
To ensure reactor security, the absence of an online connection extended downtime during summer 2022. Despite these challenges, the data acquisition (DAQ) system maintained an average efficiency of approximately 70\% throughout the data-taking period. 

At the start of physics operations, we collected data while the reactor was operating at full power (reactor-on data) for 120 days. However, an unexpected power outage caused the NEON DAQ system to be offline for 38 days during this period. The reactor was inactive from September 26, 2022, to February 22, 2023, for regular maintenance and fuel replacement, during which reactor-off data were collected for 144 days. After maintenance, the reactor resumed operation on February 22, 2023. To avoid complexities arising from changes in photon and ALP fluxes, data from ramp-down  and ramp-up periods  were excluded. Once the reactor restarted, it operated stably at full power. Data collected through June 22, 2023, added an additional 117 days of reactor-on exposure.

Various internal and external radiation peaks were used to calibrate the energy scale and resolution, following procedures similar to those adopted in the COSINE-100 experiment~\cite{COSINE-100:2024ola,COSINE-100:2024log}. Internal and external background peaks include 31\,keV from $^{121m}$Te, 39 and 67\,keV from $^{125}$I, 49\,keV from $^{210}$Pb,  295\,keV from $^{214}$Pb,  609 and 1764\,keV from $^{214}$Bi, 1461\,keV from $^{40}$K, and 2615\,keV from $^{208}$Tl. External calibrations were also performed using radioactive sources, yielding peaks at  60\,keV from $^{241}$Am and 511 and 1274\,keV from $^{22}$Na~\cite{NEON:2022hbk}. It is well known that scintillators like NaI(Tl) crystals exhibit a nonproportional relationship between energy deposition and light output~\cite{nonprop}. The nonproportional response model, characterized using the COSINE-100 NaI(Tl) crystals~\cite{COSINE-100:2024log}, was applied to correct the energy scale used in this analysis.

This analysis focused on events with energies between 3\,keV and 3000\,keV. 
The region below 3\,keV was dominated by PMT-induced noise pulses and afterpulses from energetic events such as cosmic muons, and inclusion of this region did not significantly enhance sensitivity to ALP signals. 
Therefore, we excluded it from the analysis to ensure robustness. Above 3\,keV, residual noise events were well rejected using a boosted decision tree (BDT)-based event selection algorithm~\cite{COSINE-100:2020wrv}, with no loss of efficiency. Although this analysis was not significantly affected by low-energy noise, the same data quality cuts developed for low-energy analyses--such as CE$\nu$NS and low-mass dark matter searches~\cite{Choi:2024pnp}--were applied for consistency. Data quality was monitored by evaluating event rates in the 1--3\,keV range after BDT selection. Each one-hour dataset was classified as ``good" if its event rate fell within 3\,$\sigma$ of the mean event rate distribution; otherwise, it was classified as ``bad". 
In total, this analysis utilized 1596 kg$\cdot$days of reactor-on data and 1467 kg$\cdot$days of reactor-off data.

Selected events were further categorized as single-hit or multiple-hit events. A multiple-hit event was defined as having accompanying crystal signals with more than four photoelectrons or a liquid scintillator signal exceeding 80\,keV within a 150\,ns  time coincidence window. Events that did not meet these criteria were classified as single-hit samples.

\begin{figure*}[!htb]
		\begin{center}
  \includegraphics[width=1.0\textwidth]{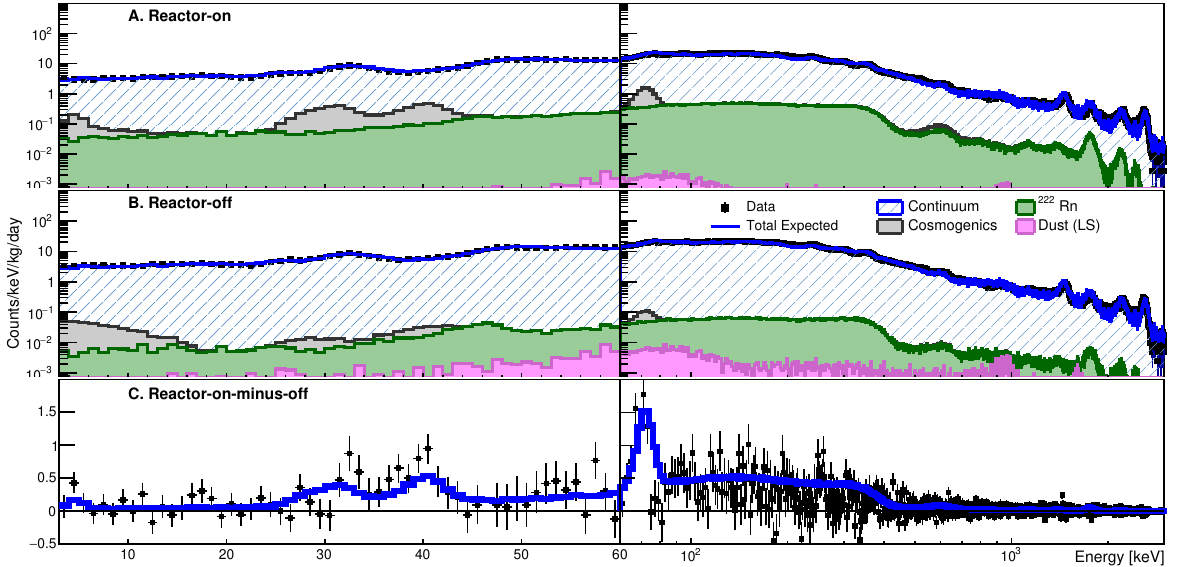} 
  \caption{
  Single-hit energy spectra of the detector-6 module. The figure shows the normalized energy spectra of single-hit events (black points) in the detector-6 module, compared with the expected background contributions (blue solid lines) for reactor-on data (A), reactor-off data (B), and the reactor-on-minus-off spectrum (C). The expected background includes time-independent continuum components and time-dependent contributions such as cosmogenic activation,  $^{222}$Rn in the calibration holes, and  $^{238}$U and $^{232}$Th from dust contamination in the liquid scintillator. For the reactor-on-minus-off spectrum (C), only time-dependent components contribute to the background.  
 }
  \label{fig:modeling_single}
	\end{center}
\end{figure*}

Most background contributions in the NaI(Tl) detectors remain stable over a 1.2\,year data acquisition period~\cite{cosinebg2}. Although the dominant $^{210}$Pb contamination has a half-life of 22.3\,years, its variation during the 1.2\,year data period is negligible. We thus define the effectively time-independent background components as the ``Continuum background", which includes internal contaminants, surface contamination, and external radiation, with a half-life equal to or greater than that of $^{210}$Pb. In addition to the continuum background, we identified a few time-dependent backgrounds. Short-lived cosmogenic contaminants in the NaI(Tl) crystals, introduced by cosmic ray exposure before installation, were characterized through dedicated analysis~\cite{COSINE-100:2019rvp}. Seasonal variations in $^{222}$Rn levels, with higher levels observed in summer due to temperature changes~\cite{Ha:2022psk}, can affect the time-dependent background. Dust contamination introduced during detector upgrades contained long-lived isotopes, which settled to the bottom of the liquid scintillator, leading to a gradual decrease in the background rate over time.

To account for time-dependent backgrounds, we divided the dataset into seven two-month periods and extracted contributions from $^{222}$Rn and dust, as detailed in the Appendix~\ref{sec_app1}. This model enables us to characterize the backgrounds observed in reactor-on (A) and reactor-off (B) periods, as examplified in the detector-6 single-hit data shown in Fig.~\ref{fig:modeling_single}. The remaining background in the reactor-on-minus-off dataset (C) is also modeled. The measured data are well described by the expected backgrounds including time-dependent background components.

\begin{figure*}[!htb]
\begin{center}
\includegraphics[width=1.0\textwidth]{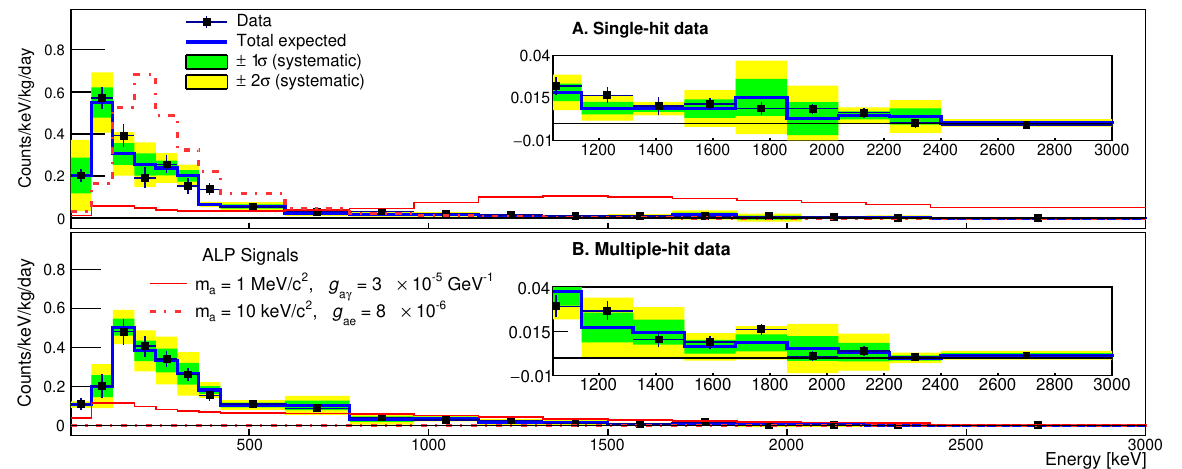} 
  \caption{ALP search data from detector-6 module.
 This figure presents the reactor-on-minus-off data spectra used for ALP signal searches  in the detector-6 module. The data is shown for both single-hit (A) and multiple-hit (B). The data points (black circles) and the expected background spectra (blue solid lines) are derived from the models presented in Fig.~\ref{fig:modeling_single}, but with different bin sizes used for this analysis.  The green and yellow bands  indicate 68\% and 95\% confidence level intervals for the background model, respectively. The inset zooms in the high-energy region for better visibility. Two benchmark ALP signals are overlaid for comparison:  m$_a$ = 10\,keV/c$^2$, $g_{ae}$ = 8$\times$10$^{-6}$ (red dashed line) and m$_a$ = 1\,MeV/c$^2$, $g_{a\gamma}$ = 3$\times$10$^{-5}$\,GeV$^{-1}$ (red solid line).   
 }
  \label{fig:ALP_input}
	\end{center}
\end{figure*}

Considering the varying event rates across different energy ranges, we employed dynamic energy bins ranging from 57\,keV (3--60\,keV) to 600\,keV (2400\,keV--3000\,keV). Figure~\ref{fig:ALP_input} presents the ALP search data from detector-6, based on the reactor-on-minus-off spectra, in which both single-hit and multiple-hit channels are used simultaneously. 

Intense photons are generated in the nuclear reactor through nuclear fission, the decay of fission products, capture processes, decay of capture products, and scattering~\cite{ROOS195998}, with the photon flux approximated from the FRJ-1 research reactor~\cite{Bechteler1984}. 
We consider a generic model in which an ALP couples to either photons or electrons. For photon coupling, ALPs can be produced through the Primakoff process ($\gamma + A \rightarrow a + A$), where a photon ($\gamma$) interacts with a nucleus ($N$) to produce an axion ($a$)~\cite{PhysRev.81.899}. Detection occurs through two-photon decay ($a\rightarrow \gamma\gamma$) or the inverse Primakoff process, with rates depending on the axion-photon coupling constant ($g_{a\gamma}$). 
For electron coupling, ALPs can occur through the Compton-like process ($\gamma+e^- \rightarrow a+e^-$) and be detected through electron-positron pair production ($a \rightarrow e^-e^+$), axio-electric absorption ($a+e^- +A \rightarrow e^- + A $), or inverse Compton-like process  ($a+e^- \rightarrow \gamma+e^-$). The rate of these processes depends on the strength of the axion-electron coupling constant ($g_{ae}$). 
To model detector responses, we employ Geant4-based simulations. 
Two benchmark ALP signals, m$_a$ = 1\,MeV/c$^2$, $g_{a\gamma}$ = 3$\times$10$^{-5}$\,GeV$^{-1}$ for axion-photon coupling and  m$_a$ = 10\,keV/c$^2$, $g_{ae}$ = 8$\times$10$^{-6}$ for  axion-electron coupling,  are compared to the measured spectra in Fig.~\ref{fig:ALP_input}. 
For ALP-electron couplings, signals typically deposit energy within a single crystal, while ALP-photon interactions can produce high-energy photons that Compton scatter across multiple detectors, leading to multiple-hit events.
We do not consider the ALP production through the nuclear de-excitation, as studied by the TEXONO experiment~\cite{TEXONO:2006spf}. 
Further details on ALP signal generation in the NEON detector are provided in Appendix~\ref{sec_app2}. 

Several sources of systematic uncertainty are included in our modeling of the reactor-on-minus-off spectra. 
These include potential variations in detector responses between the reactor-on period and the reactor-off periods. 
The largest systematic uncertainties are associated with the time-dependent background modeling of $^{222}$Rn variation, as well as uranium and thorium concentrations in the dust. 
Additionally, possible contamination of $^{222}$Rn into the liquid scintillator and different locations of the dust also contribute to systematic uncertainties. 
Figure~\ref{fig:ALP_input} indicates systematic uncertainty bands for the expected background. While the reactor-on-minus-off background rate is allowed to vary by approximately 20\% to account for these effects, the resulting impact on the exclusion limit is modest: the sensitivity to the ALP coupling constant changes by less than 3\%, as the signal rate scales with the fourth power of the coupling constant. 
Furthermore, we account for a 10\% uncertainty in the reactor photon flux, originating from differences between the FRJ-1 research reactor model and the commercial Hanbin reactor, for which no detailed photon spectrum is publicly available. This 10\% variation leads to an additional 2.4\% change in the derived axion-photon and axion-electron coupling limits.

The NEON data are fitted for each ALP mass and interaction type. 
We use various simulated ALP signals to evaluate their potential contributions to the measured energy spectra of the reactor-on-minus-off data (shown in Fig.~\ref{fig:ALP_input}). A $\chi^2$ fit is applied to the measured spectra  (both single-hit and the multiple-hit channels) between 3 and 3,000\,keV for each ALP signal and various ALP masses. Each crystal and channel is fitted with a crystal-channel specific background model and a crystal-channel correlated ALP signal. The combined fit is achieved by summing the $\chi^2$ values from the all crystals and channels. 
No statistically significant excess of events was found for any of the considered ALP signals. The posterior probabilities of the signals are consistent with zero in all cases, and 95\% confidence level limits are determined.

\begin{figure*}[!htb]
		\begin{center}
			\begin{tabular}{c c}
      \includegraphics[width=0.5\textwidth]{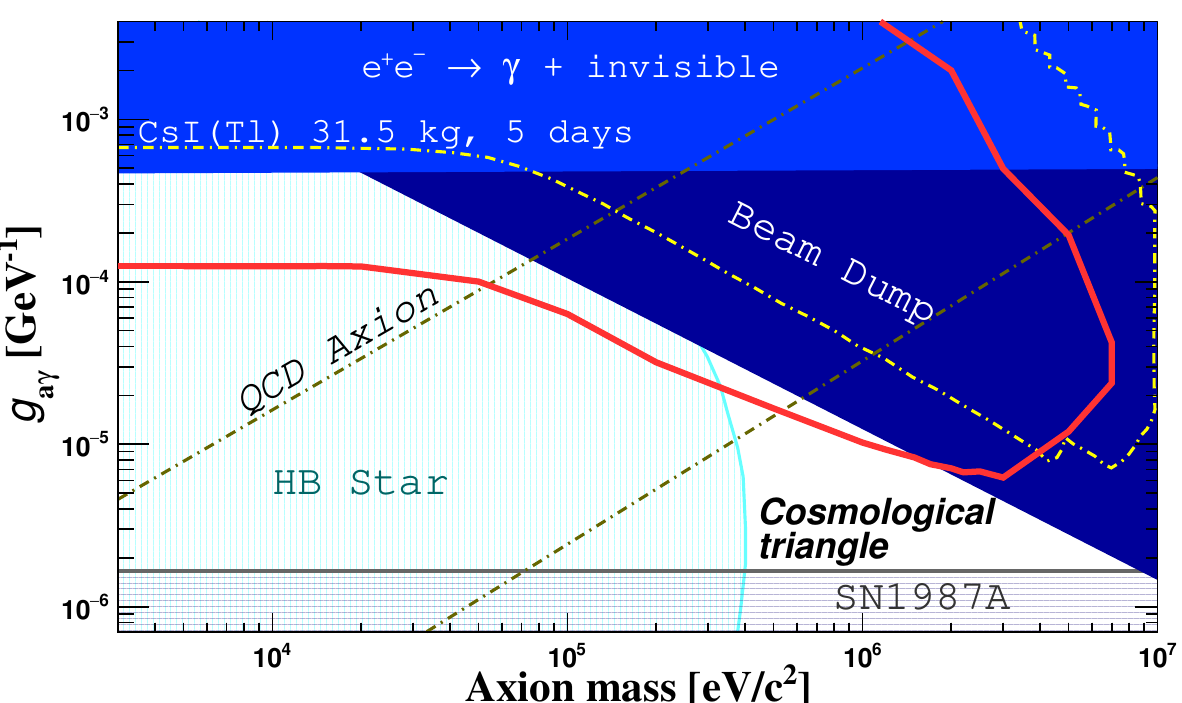} &
      \includegraphics[width=0.5\textwidth]{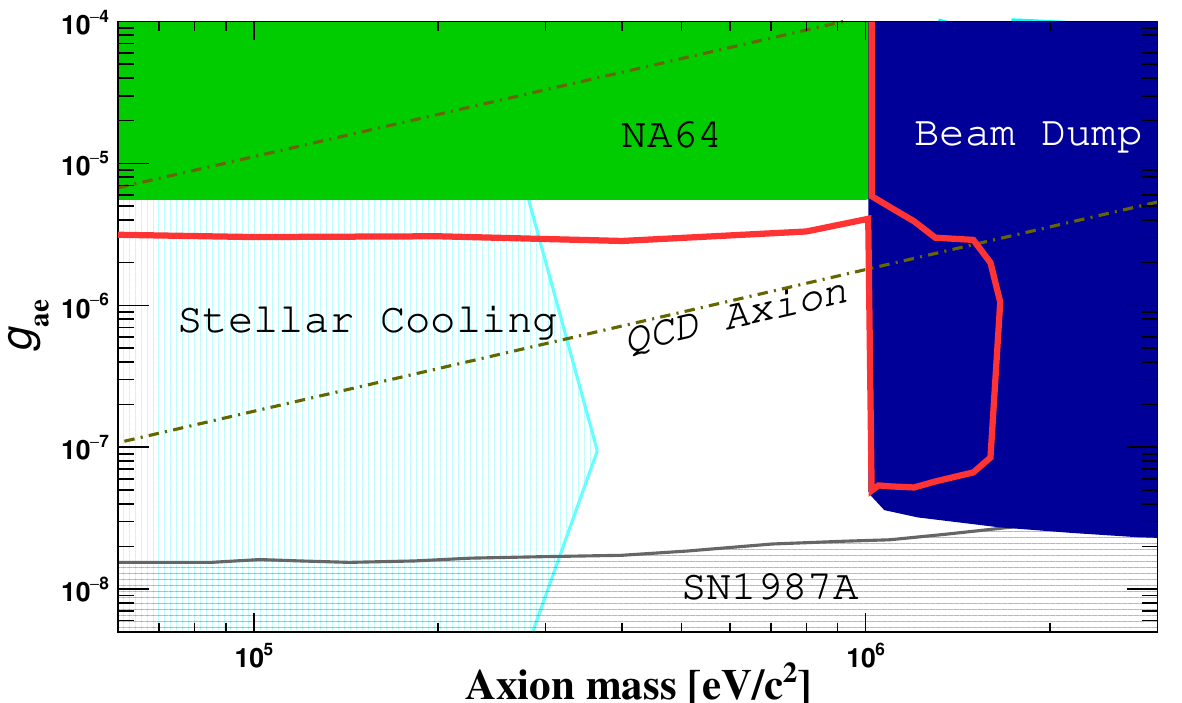}\\
			(a) Axion-photon coupling & (b) Axion-electron coupling \\
    	\end{tabular}
			\caption{Exclusion limits of the ALP interactions. The observed 95\% confidence level exclusion limit (red solid line) derived from NEON data for the axion-photon (a) and the axion-electron (b) are  compared with limits from beam dump experiments~\cite{Jaeckel:2015jla,Blumlein:1991xh,Bjorken:1988as,Bechis:1979kp,Riordan:1987aw}, SN1987A~\cite{Payez:2014xsa,Jaeckel:2017tud, Lucente:2021hbp}, $e^{+}e^{-}\rightarrow\gamma+$invisible states~\cite{Dolan:2017osp}, HB stars cooling arguments~\cite{Carenza:2020zil}, stellar cooling constraints~\cite{Hardy:2016kme}, the NA64 missing energy search~\cite{Gninenko:2017yus}, and CsI(Tl) exposure in a nuclear reactor~\cite{Sahoo:2024zee}. The QCD axion model parameter spaces of the KSVZ~\cite{DiLuzio:2020wdo} and DFSZ(I)~\cite{DiLuzio:2020wdo} benchmark scenarios for  the axion-photon coupling and  axion-electron coupling, respectively, are indicated by the gray dashed lines. 
    }
  \label{figure:garr}
	\end{center}
\end{figure*}

Figure~\ref{figure:garr} (a) presents the 95\% confidence level exclusion limit derived from NEON data for ALPs coupled solely to photons. This limit is shown in the two-dimensional parameter space of m$_a$--$g_{a\gamma}$. For ALP masses below  20\,keV/c$^2$, the dominant contribution arises from the scattering process via the inverse Primakoff process. At higher ALP masses, the limit is set by the $a\rightarrow\gamma\gamma$ decay process. Since this decay can occur within the 23.7\,m flight path, limits are considered for both lower and higher $g_{a\gamma}$ values. For ALP masses above 3\,MeV/c$^2$, sensitivity declines  due to detector saturation effects and the decreasing reactor photon flux at higher energies. However, signatures of Compton scattering could still allow searches for higher-mass ALPs. In this process, a high-energy photon from the ALP interaction deposits a lower-energy electron or photon within the detectable range.  
Future improvements, such as reconstructing saturated events--similar to techniques employed in the COSINE-100 experiment for boosted dark matter searches~\cite{COSINE-100:2023tcq}--could enhance sensitivity to higher-mass ALPs.

The exclusion limits shown in Fig.~\ref{figure:garr} (a) extend to previously unexplored regions of ALP parameter space, surpassing existing constraints from beam dump experiments and astrophysical and cosmological limits as adapted from Refs.~\cite{Fortin:2021cog,Batell:2022xau}. Notably, this study starts  to probe the ``cosmological triangle'', a previously unconstrained region between beam dump experiments and astrophysical bounds. A small remaining region of the KSVZ QCD axion model parameter space~\cite{DiLuzio:2020wdo}, corresponding to axion masses of a few hundred keV/c$^2$, is partially ruled out. The exclusion limit reaches lower $g_{a\gamma}$ values, down to 6.24$\times$10$^{-6}$\,GeV$^{-1}$ for m$_a$ = 3.0\,MeV/c$^2$.  
 Compared to a recent reactor-based ALP search using CsI(Tl) crystals~\cite{Sahoo:2024zee}, the NEON experiment significantly improved the lower bound on $g_{a\gamma}$, benefiting from a larger exposure and lower background levels in NaI(Tl) crystals. However, the greater distance from the reactor core to the NEON detector results in a reduced upper bound. This work provides new experimental constraints from a direct ALP search with reduced model dependence. 

 Figure~\ref{figure:garr} (b) displays the 95\% confidence level exclusion limit for ALPs coupled purely to electrons, presented in the m$_a$--$g_{ae}$ parameter space. For ALP masses below 1.02\,MeV/c$^2$, the limit is primarily set by scattering processes via the inverse Compton-like process and axio-electric absorption. For higher ALP masses (m$_a>$ 1.0\,MeV/c$^2$), the limit is dominated by the $a\rightarrow e^+e^-$ decay process, which has a kinematic threshold of m$_a >$ 2m$_e$ = 1.02\,MeV/c$^2$ (where m$_e$ is the electron mass). Similar to the ALP-photon case, limits are considered for both upper and lower bounds due to potential ALP decay during flight.  

The NEON data also explore previously examined regions constrained by stellar cooling arguments~\cite{Raffelt:2006cw} for axion masses below 300\,keV/c$^2$, where environmental effects could allow circumvention of these limits~\cite{AristizabalSierra:2020rom}.
In the mass range of 300\,keV/c$^2$ and 1.02\,MeV/c$^2$, scattering processes probe coupling values down to  $g_{ae}$ about 3$\times$10$^{-6}$, which were previously unexplored by direct searches or astrophysical and cosmological considerations. This limit extends into regions predicted by the DFSZ-I QCD axion model~\cite{DiLuzio:2020wdo}. For ALP masses above the kinematic limit for $a\rightarrow e^+ e^-$ (m$_a >$1.02\,MeV/c$^2$), the NEON data compete with limits from beam dump experiments~\cite{Bjorken:1988as,Bechis:1979kp,Riordan:1987aw}. The exclusion limit reaches lower $g_{ae}$ values, down to 4.95$\times$10$^{-8}$ for m$_a$ = 1.02\,MeV/c$^2$. 
The NEON search for axion-electron coupling is currently limited to ALP masses below 1.6\,MeV/c$^2$ due to the 3\,MeV dynamic range of the analysis. Similar to the ALP-photon case, reconstructing saturated events above 3\,MeV energies could extend the search to higher ALP masses, as demonstrated in  Refs.~\cite{AristizabalSierra:2020rom,Sahoo:2024zee}.

This study reports a direct search for axion-like particles (ALPs) using the NEON experiment.  Leveraging 16.7\,kg of NaI(Tl) crystals located 23.7\,m from a 2.8\,GW thermal power reactor core, NEON has set new exclusion limits for ALPs coupling to photons and electrons. These results probe previously inaccessible regions of ALP parameter space, particularly axion masses around  1\,MeV/c$^2$, for both axion-photon and axion-electron couplings. Future improvements, such as increased data exposure and advanced  reconstruction of saturated events above 3\,MeV, will further enhance  NEON's sensitivity to ALP searches. 

\acknowledgments
We thank the Korea Hydro and Nuclear Power (KHNP) company for the help and support provided by the staff members of the Safety and Engineering Support Team of Hanbit Nuclear Power Plant 3 and the IBS Research Solution Center (RSC) for providing high performance computing resources. 
This work is supported by the Institute for Basic Science (IBS) under Project Code IBS-R016-A1 and the National Research Foundation (NRF) grant funded by the Korean government (MSIT) (NRF-2021R1A2C1013761 and NRF-2021R1A2C3010989), Republic of Korea.
GK and JN are supported by Fermi Research Alliance, LLC under Contract DEAC02-07CH11359 with the U.S. Department of Energy and the Kavli Institute for Cosmological Physics at the University of Chicago through an endowment from the Kavli Foundation and its founder Fred Kavli.
\providecommand{\href}[2]{#2}\begingroup\raggedright\endgroup

\bibliographystyle{PRTitle}

\clearpage
\renewcommand{\thefigure}{A\arabic{figure}}
\renewcommand{\thetable}{A\arabic{table}}
\renewcommand{\theequation}{A\arabic{equation}}
\setcounter{figure}{0}
\setcounter{table}{0}
\setcounter{equation}{0}
\setcounter{section}{0}

\section{Appendix}\label{sec8}

\subsection{Time-dependent background model}\label{sec_app1}
\begin{figure}[!htb]
		\begin{center}
      \includegraphics[width=1.0\columnwidth]{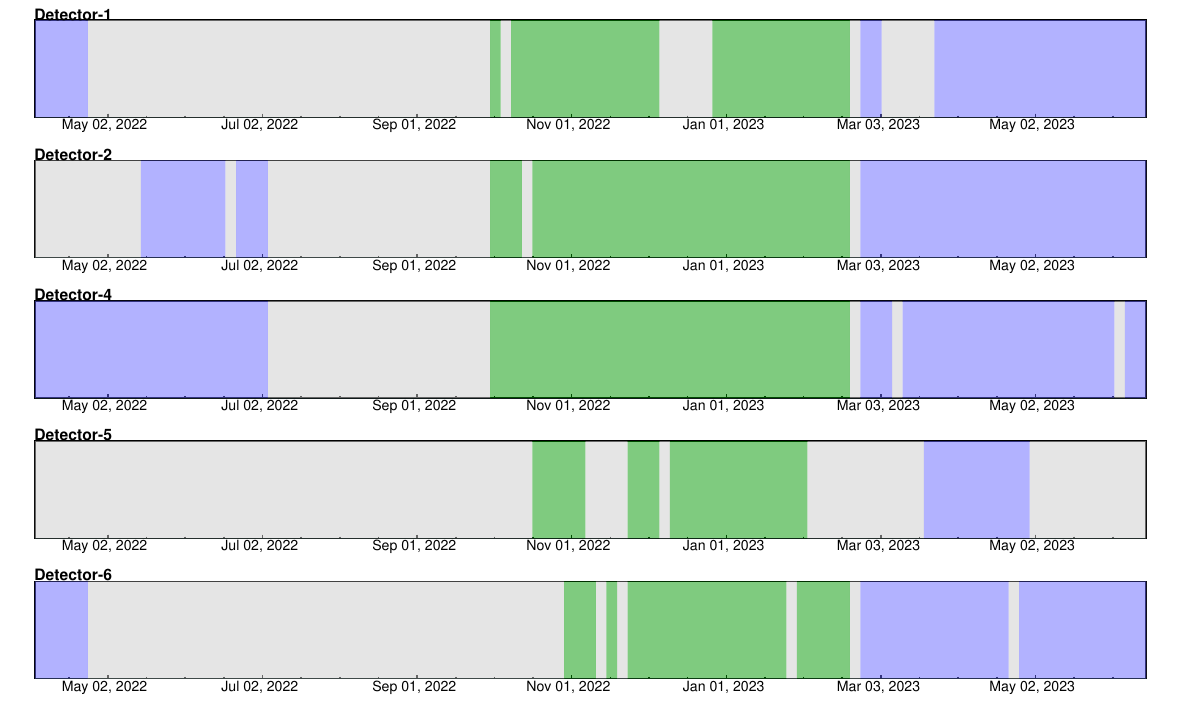} 
  \caption{Good quality data in the time domain. The figure shows good quality data as a function of time for each detector module. Reactor-on data is depicted in blue, reactor-off data in green, and periods of DAQ downtime or poor-quality data are indicated in gray. Data from detector-3 were excluded from the analysis due to persistent low-energy noise. 
    }
  \label{fig:gooddata}
	\end{center}
\end{figure}

The cosmogenic contributions to the NaI(Tl) crystal detectors were extensively studied in the the ANAIS~\cite{Amare_2015,S0217751X18430066} and COSINE~\cite{COSINE-100:2019rvp} experiments. These contributions include isotopes such as $^{125}$I, $^{121}$Te, $^{123m}$Te,$^{125m}$Te,$^{127m}$Te, $^{113}$Sn, $^{22}$Na and $^{3}$H, which primarily affect energy ranges below 100\,keV. Long-lived isotopes such as $^{22}$Na (2.6 years) and $^{3}$H (12.3 years) persist across all detectors. In contrast, short-lived isotopes (half-lives$<$ 1 year) decayed following the initial installation in December 2020. Notably, the initial data from detector-5 and detector-6 were significantly affected by short-lived cosmogenic isotopes, as these crystals were replaced shortly before the start of physics operations.  Detector-1 exhibited  relatively high cosmogenic activation due to approximately one year of additional muon exposure during R\&D on detector encapsulation. 
Based on exposure times shown in Fig.~\ref{fig:gooddata} and the half-life of each isotope, we modeled the cosmogenic contributions for each detector module.

Another significant source of time-dependent background is the seasonal variation of $^{222}$Rn, which has been reported to be higher during the summer and lower during the winter~\cite{LI2006101,Ha:2022psk}. The NEOS experiment~\cite{NEOS:2016wee} measured $^{222}$Rn levels in the tendon gallery using a Radon eye device, confirming this seasonal variation. Initially, the NEON experiment did not include a $^{222}$Rn monitoring device. However, in December 2023, a Radon eye was installed to monitor the $^{222}$Rn levels in the tendon gallery.

\begin{figure*}[!htb]
\includegraphics[width=0.32\textwidth]{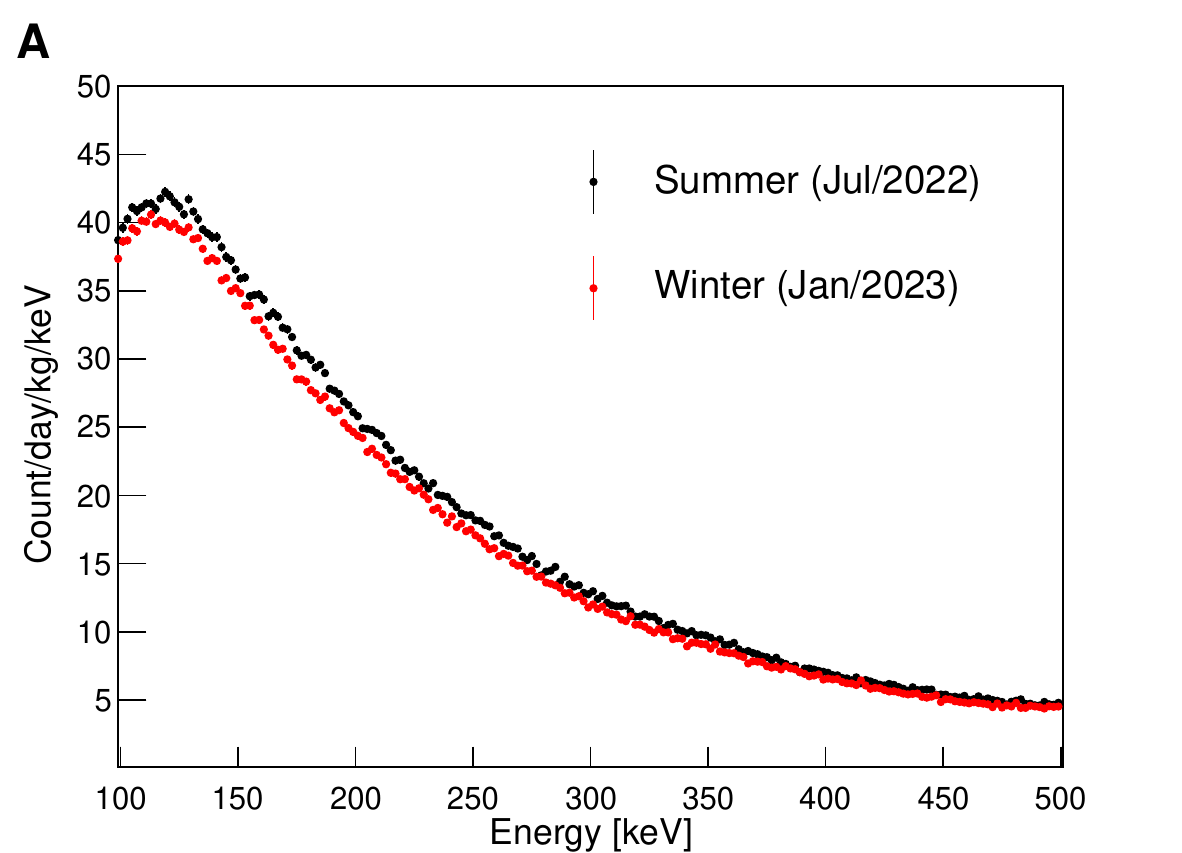} 
 \includegraphics[width=0.32\textwidth]{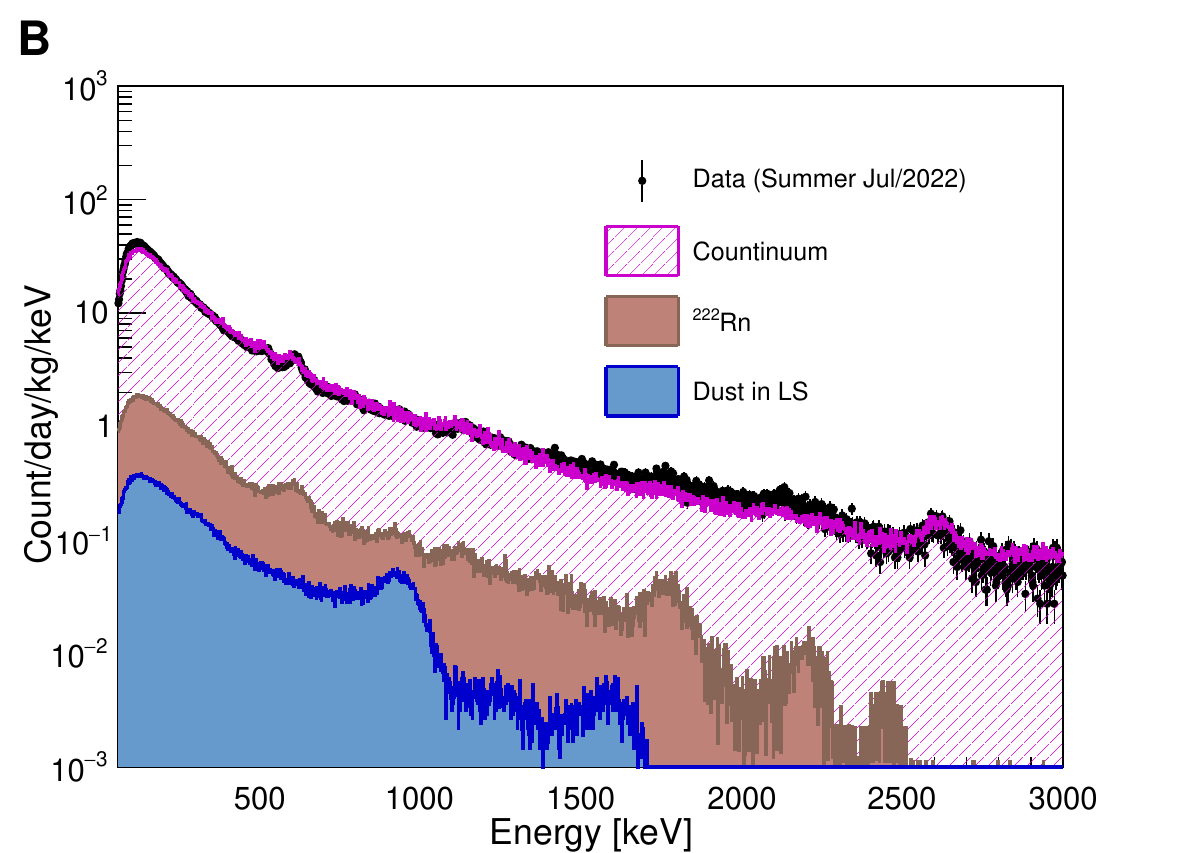} 
\includegraphics[width=0.32\textwidth]{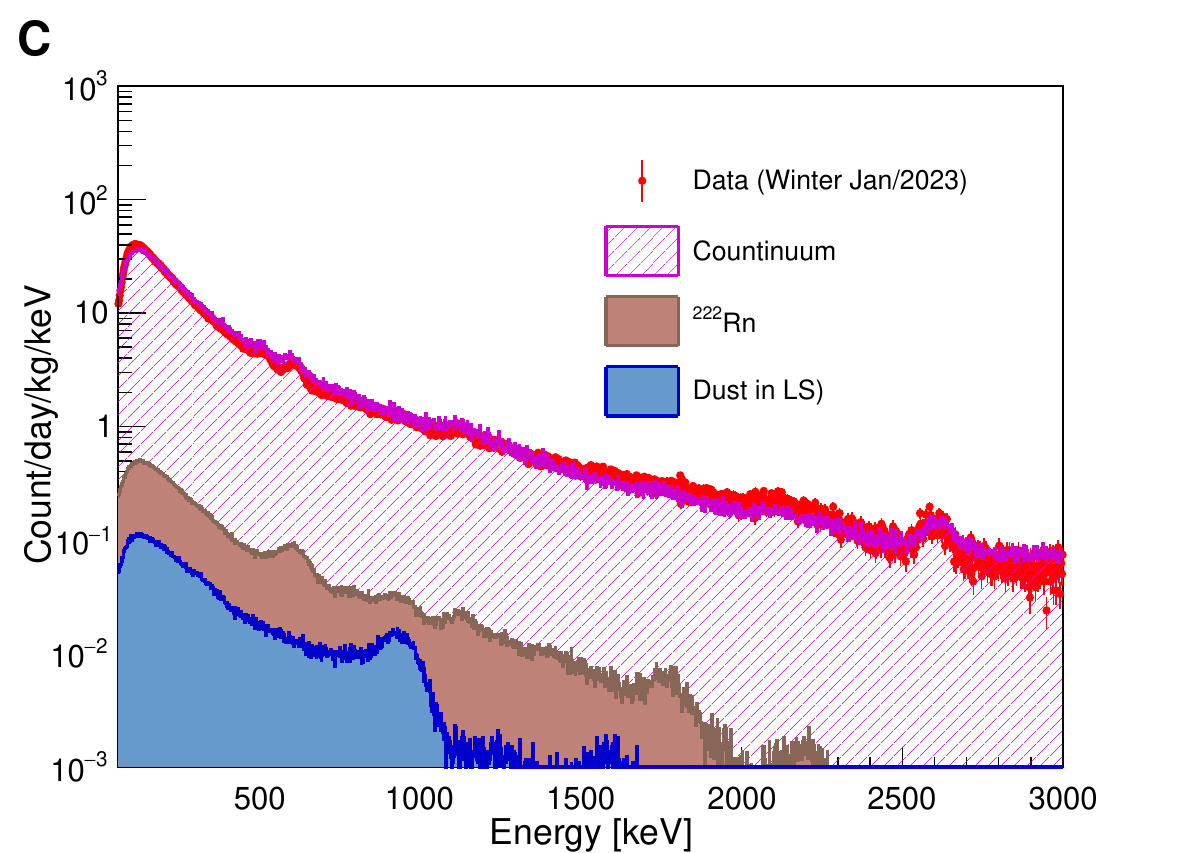} 
  \caption{ 
  Time-dependent background models for multiple-hit events in detector-6. Panel (A) shows a noticeable difference in event rates between the summer (15 June 2022 – 15 August 2022) and winter (1 January 2023 – 28 February 2023) seasons, attributed to the seasonal variation of $^{222}$Rn. Panels (B) and (C) illustrate data from summer and winter, respectively, modeled with the time-independent continuum background and time-dependent contributions from $^{222}$Rn and liquid scintillator dust, quantifying the amounts of these components.
  }
  \label{fig:Rn222change}
\end{figure*}

The NEON detector includes two calibration holes extending from the top of the shield to the vicinity of the crystal modules~\cite{NEON:2022hbk}. These calibration holes were exposed to the same levels of $^{222}$Rn as the experimental tunnel. As observed in the NEOS experiment, these seasonal $^{222}$Rn variations likely influenced the background levels recorded in the NEON experiment. Simulated spectra for $^{222}$Rn in the calibration holes revealed that the largest contributions occur within the 100--500\,keV range for multiple-hit events (see Fig.~\ref{fig:Rn222change}(A)). Summer data showed significantly elevated rates compared to winter data, predominantly due to seasonal variations in $^{222}$Rn. Because  most reactor-off data were collected during winter (see Fig.~\ref{fig:gooddata}), seasonal $^{222}$Rn fluctuations led to higher background levels in reactor-on data.

In addition to radon, another significant time-dependent background component is dust contamination in the liquid scintillator.
The environmental conditions in the tendon gallery contain a significant amount of dust. The only way to minimize dust contamination was to complete all installations quickly and seal the detector system. However, due to unstable detector conditions observed during engineering run~\cite{NEON:2024Upgrade}, approximately one year was spent upgrading the detector encapsulation and conducting various tests with the detector-1 module. During this period, the liquid scintillator was drained and refilled multiple times, leading to dust contamination. At the beginning of the experiment, the liquid scintillator was refilled, and dust particles may have been suspended throughout the scintillator. Over time, these particles settled at the bottom of the scintillator, leading to  a gradual reduction in background rates.

\begin{figure}[!htb]
      \includegraphics[width=1.0\columnwidth]{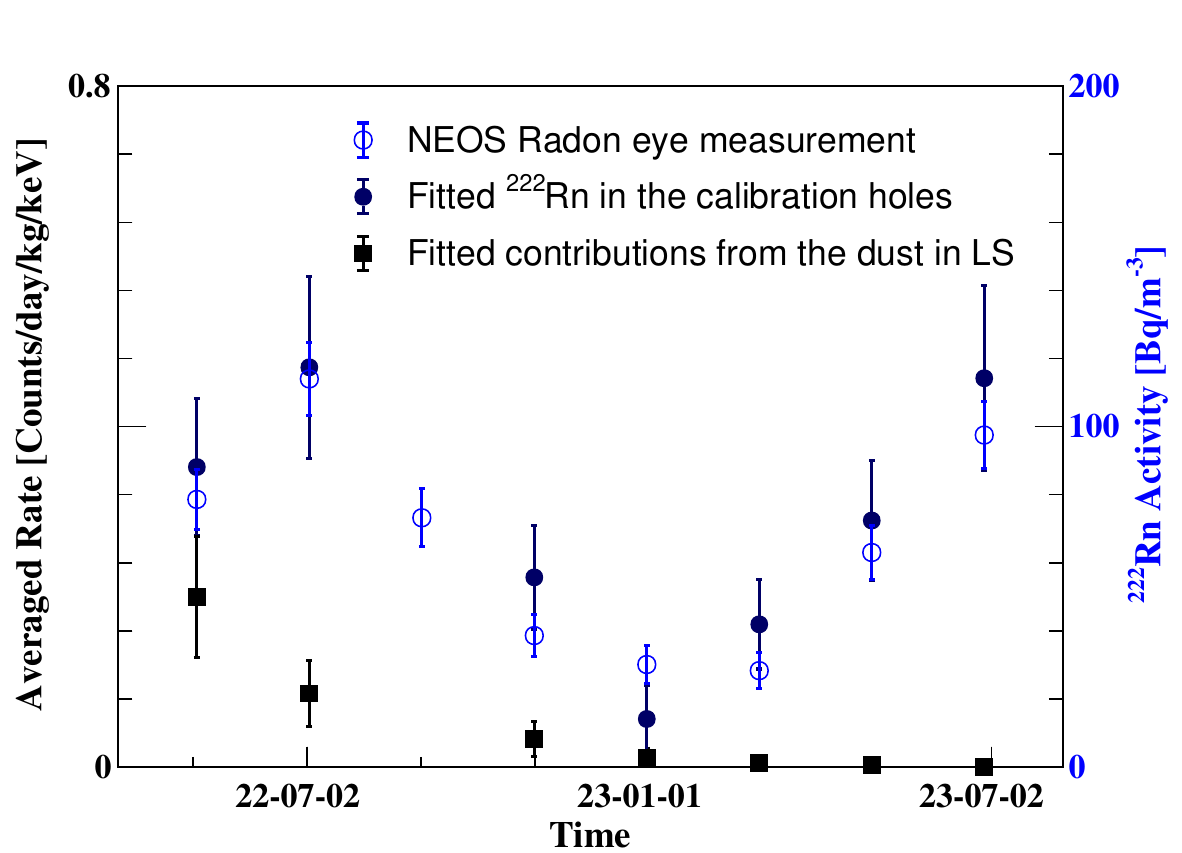} 
  \caption{ 
  Time-dependent background contributions from  $^{222}\mathrm{Rn}$  and liquid scintillator dust contamination. The fitted rates for multiple-hit events in detector-6, averaged over energies between 100\,keV and 500\,keV, are presented (left axis). Contributions include  $^{222}\mathrm{Rn}$  in the calibration holes (dark blue filled circles) and liquid scintillator dust (black filled squares). Measured  $^{222}\mathrm{Rn}$  activities (right axis) from the NEOS tunnel using a Radon Eye device (blue open circles) are also shown, corresponding to the same seasons, although these measurements were conducted in different years and in the tendon gallery of reactor unit-5 (while NEON is located at reactor unit-6). Despite differences in location and year, the seasonal variations of  $^{222}\mathrm{Rn}$  show excellent agreement between the NEOS Radon Eye measurements and the NEON data.
    }
  \label{fig:rn222td}
\end{figure}

To model time-dependent background contributions, the data were divided into seven time periods, each spanning two months. 
Each dataset was modeled using known NaI(Tl) background components studied in the COSINE-100 experiment~\cite{COSINE-100:2024ola} as shown in Fig.~\ref{fig:Rn222change}. In this process, contributions from $^{222}$Rn and dust were extracted as summarized in Fig.~\ref{fig:rn222td}. The extracted $^{222}$Rn contamination levels from the NEON data showed excellent agreement with NEOS measurements using the Radon eye device.

\begin{figure}[!htb]
		\begin{center}
  \includegraphics[width=1.0\columnwidth]{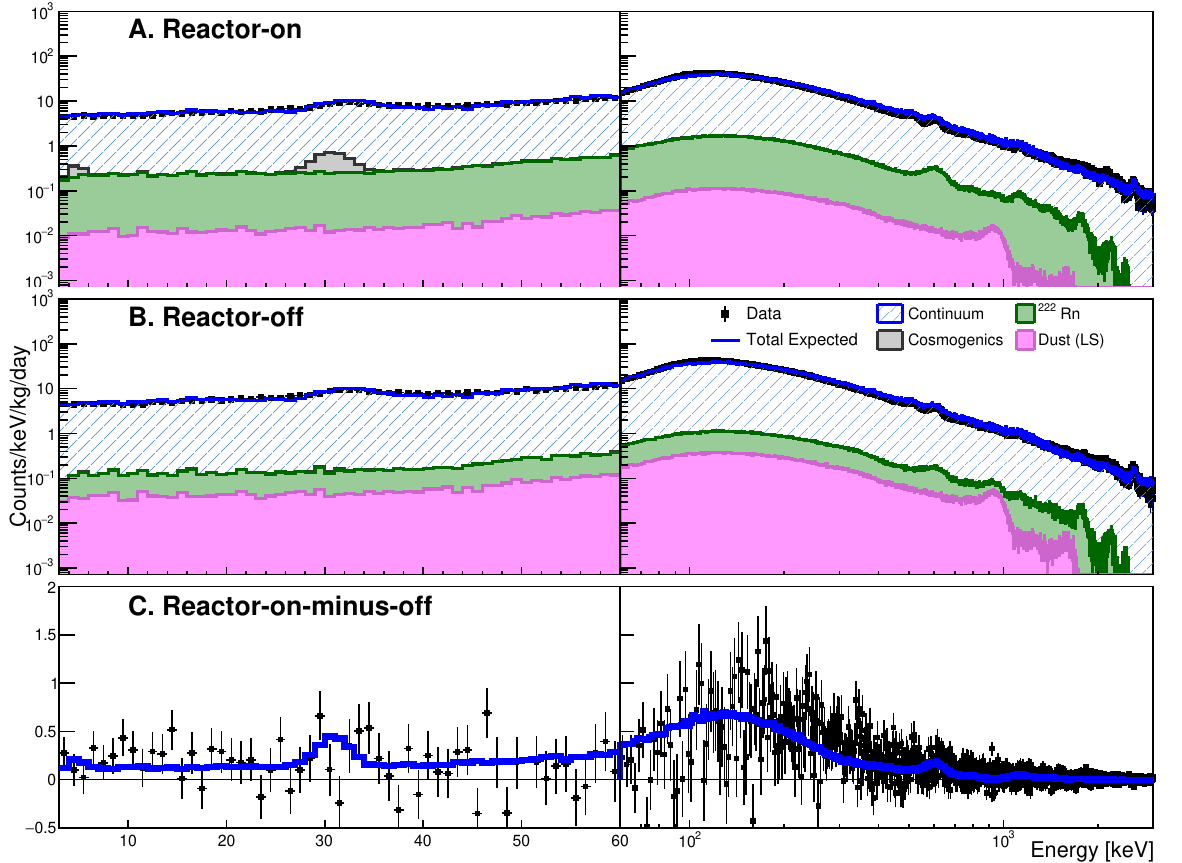}
  \caption{
  Multiple-hit energy spectra of the detector-6 module. The figure shows the normalized energy spectra of single-hit events (black points) in the detector-6 module, compared with the expected background contributions (blue solid lines) for reactor-on data (A), reactor-off data (B), and the reactor-on-minus-off spectrum (C). The expected background includes time-independent continuum components and time-dependent contributions such as cosmogenic activation,  $^{222}$Rn in the calibration holes, and  $^{238}$U and $^{232}$Th from dust contamination in the liquid scintillator. For the reactor-on-minus-off spectrum (C), only time-dependent components contribute to the background.  
    }
  \label{fig:modeling_multiple}
	\end{center}
\end{figure}

Based on the time-dependent background analysis for each crystal, background contributions for reactor-on (A) and reactor-off (B) dataset were modeled, as shown in Fig.~\ref{fig:modeling_multiple} for multiple-hit data. Background components were categorized into continuum, cosmogenic, $^{222}$Rn, and dust contributions. 
The remaining backgrounds in the reactor-on-minus-off dataset were modeled using the time-dependent backgrounds of the cosmogenic isotopes, $^{222}$Rn, and dust, as shown in Fig.~\ref{fig:modeling_multiple}(C). The measured data align well with the expected background models. Similar examples for single-hit data were included in the main text (Fig.~\ref{fig:modeling_single}). 

\subsection{ALP Signal Generation}\label{sec_app2}

Nuclear reactor cores produce a vast number of photons, which can scatter off fuel materials within the reactor tank to generate ALPs~\cite{ROOS195998}.
The photon flux is approximated using the FRJ-1 research reactor model~\cite{Bechteler1984}, expressed as: 
\begin{eqnarray}
  \frac{d\Phi_{\gamma}}{dE_{\gamma}} = \frac{5.8 \times 10^{17}}{[{\rm MeV}]\cdot[{\rm sec}]}\left(\frac{P}{[{\rm MW}]}\right)e^{-1.1E_{\gamma}/[{\rm MeV}]},
\end{eqnarray}
where $P$ is the thermal power and $E_{\gamma}$ is the photon energy. A systematic uncertainty of up to 10\% in the photon flux was considered; however, its impact on the derived coupling constants is negligible~\cite{AristizabalSierra:2020rom}. 

We consider a generic model where ALPs couple to photons ($g_{a\gamma}$) or  electrons ($g_{ae}$)~\cite{PhysRevD.81.123530, Cicoli2012}. 
ALPs can be produced via the Primakoff process ($\gamma + A\rightarrow a + A$)~\cite{PhysRev.81.899} and the Compton-like process ($\gamma + e^{-} \rightarrow a + e^{-}$)~\cite{AristizabalSierra:2020rom}. 
After production in the reactor core, ALPs propagate through shielding materials, either decaying in flight or reaching the detector. 
The ALP flux at the detector is described as:
\begin{align}
  \frac{d\Phi^{P}_{a}}{dE_{a}} = & P_{\rm surv} \int^{E_{\gamma, max}}_{E_{\gamma,min}} \frac{1}{\sigma_{SM}+\sigma^{P}_{P(C)}}\frac{d\sigma^{p}_{P(C)}} {dE_{a}}\left(E_{\gamma},E_{a}\right) \nonumber \\
	                               & \times \frac{d\Phi_{\gamma}}{dE_{\gamma}}dE_{\gamma},
  \label{eq:prdalpflux}
\end{align}
 where $\sigma_{SM}$ is the total photon scattering cross-section against core material, referenced from the Photon Cross Sections Database~\cite{XCOM}, $E_{a}$ is the energy of the ALP, and $\sigma^{P}_{P(C)}$ is the production cross-section for the Primakoff process (Compton-like process)~\cite{AristizabalSierra:2020rom}. 

The ALP survival probability $P_{\rm surv}$ to the detector is given by~\cite{PhysRevLett.124.211804},
\begin{eqnarray}
  P_{\rm surv} = e^{-LE_{a}/p_{a}\tau},
  \label{eq:psurv}
\end{eqnarray}
where $L$ is the distance from the reactor core to the detector, $p_{a}$ is the ALP momentum, and $\tau$ is the ALP lifetime. 
The lifetime is determined by the decay widths: 
\begin{eqnarray}
 \Gamma(a\rightarrow \gamma\gamma) = \frac{g^{2}_{a\gamma} m^{3}_{a}}{64\pi} \\
 \Gamma(a\rightarrow e^{+}e^{-}) = \frac{g^{2}_{ae}m_{a}}{8\pi}\sqrt{1-4\frac{m^{2}_{e}}{m^{2}_{a}}}, 
\end{eqnarray}
where $m_{a}$ and $m_{e}$ are the ALP and electron masses, respectively.

In case of non-zero $g_{a\gamma}$, the ALPs could be detected through the inverse Primakoff process ($a + A\rightarrow \gamma + A$) or two photon pair decay  ($a \rightarrow \gamma\gamma$) in the detector material. The expected signal rate for this process is given by:
\begin{eqnarray}
\frac{dN_{a\gamma}}{dE_{a}} =   \frac{N_{target}}{4\pi L^{2}} \sigma^{\rm P}_{\rm D} \frac{d\Phi^{P}_{a}}{dE_{a}} +\frac{A}{4\pi L^{2}}\frac{d\Phi^{P}_{a}}{dE_{a}}P_{\rm decay},
\end{eqnarray}
where $N_{target}$ is  the number of target nuclei and $\sigma^{\rm P}_{\rm D}$ is the total inverse-Primakoff scattering cross section. 
$A$ is the detector transverse area, and $P_{\rm decay}$ is the probability of decay within the detector: 
\begin{eqnarray}
  P_{\rm decay} = 1 - e^{-L_{det}E_{a}/|p_{a}|\tau}, 
  \label{eq:pdecay}
\end{eqnarray}
where $L_{det}$ is the detector length in ALP flight direction. 

In case of non-zero $g_{ae}$, the ALPs could be detected through the inverse-Compton-like process ($a+e^{-}\rightarrow\gamma+e^{-}$), Axio-electric absorption ($a+e^{-}+A\rightarrow e^{-}+A$)~\cite{Avignone:1986vm}, or electron-positron pair decay ($a \rightarrow e^{+}e^{-}$). The expected signal rate for this is given by:
\begin{align}
  \frac{dN^{C}_{C}}{dE_{a}} = & \frac{N_{target}}{4\pi L^{2}} \sigma^{ C}_{ D} \frac{d\Phi^{ C}_{a}}{dE_{a}} + \frac{N_{target}}{4\pi L^{2}} \sigma^{A}_{D} \frac{d\Phi^{\rm C}_{a}}{dE_{a}} \nonumber \\ 
	                            & +\frac{A}{4\pi L^{2}}\frac{d\Phi^{P}_{a}}{dE_{a}}P_{\rm decay},
\end{align}
where $\sigma^{C}_{D}$ and $\sigma^{A}_{D}$ are the inverse-Compton-like process cross section and the Axio-electron cross section, respectively~\cite{Bellini2008,AristizabalSierra:2020rom}.

\begin{figure}[!htb]
  \includegraphics[width=0.49\textwidth]{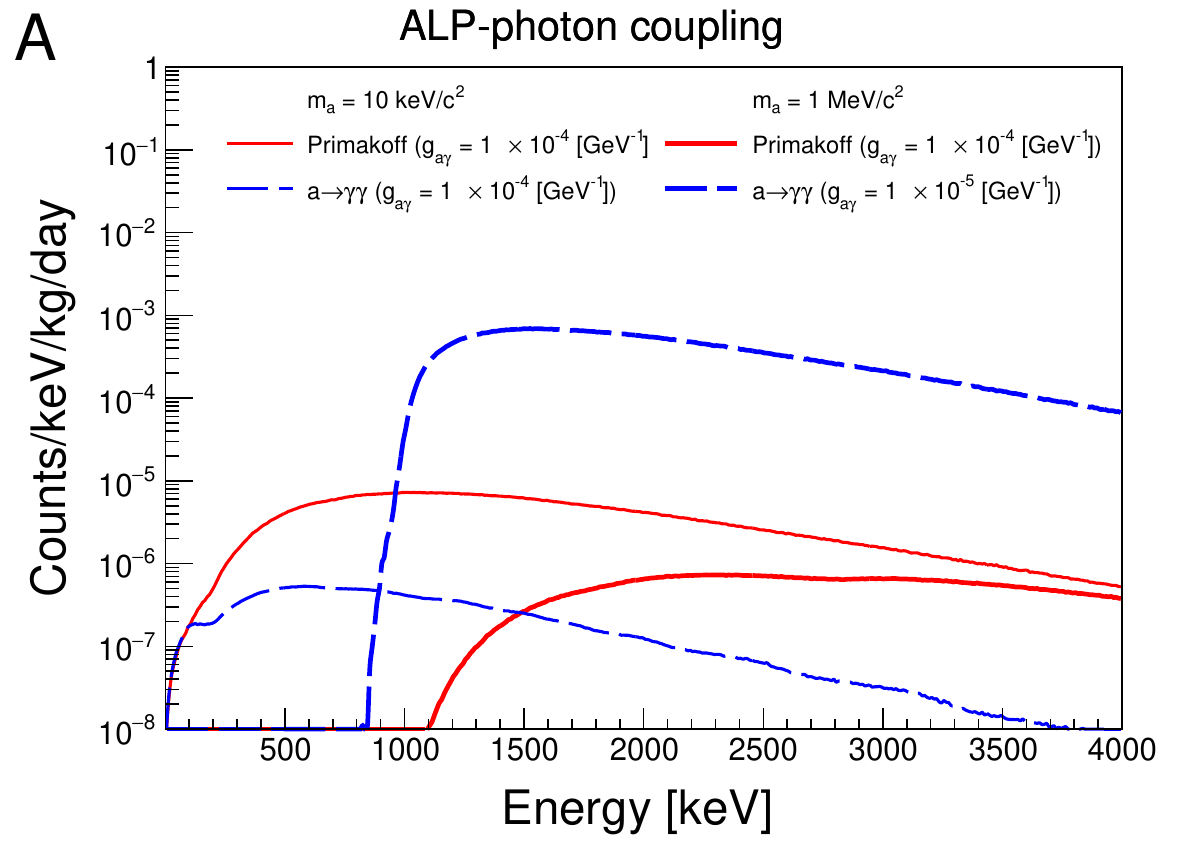}
  \includegraphics[width=0.49\textwidth]{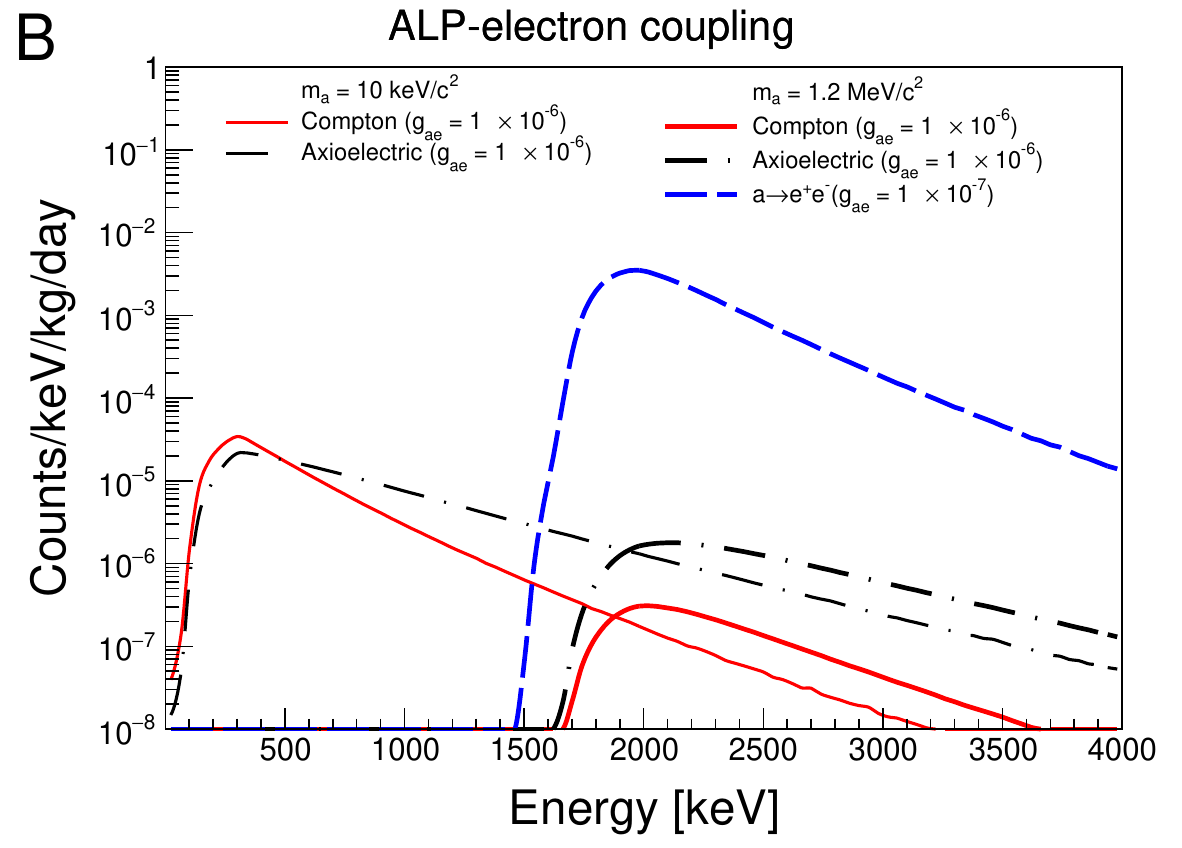} 
  \caption{Expected ALP events rates for each detection process. The scatter and decay rates from axion-photon coupling (A) and axion-electron coupling (B) in the NaI(Tl) crystal are shown for selected values of $g_{a\gamma}$ and $g_{ae}$, corresponding to ALP masses of 10\,keV/c$^2$, 1\,MeV/c$^2$, and 1.2\,MeV/c$^2$. The total energy of the produced standard model particles is presented. 
  }
  \label{figure:ALPEventrate}
\end{figure}

Figure~\ref{figure:ALPEventrate} shows the expected event rates in the NaI(Tl) crystals for various detection processes.  
These energy spectra account for the total energy of produced standard model particles such as electrons, positrons, and photons in the NaI(Tl) crystals during ALP interactions.   

\begin{figure}[!htb]
\centering
      \includegraphics[width=0.49\textwidth]{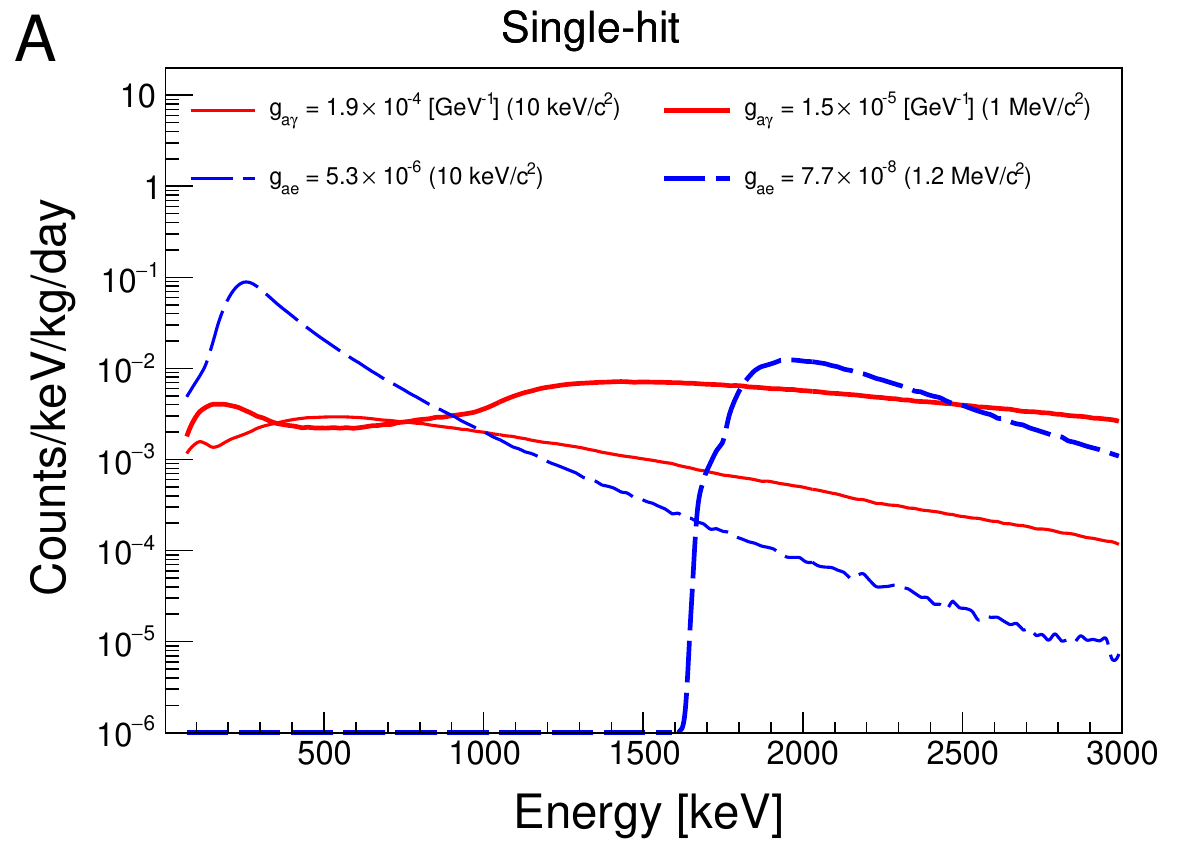} 
      \includegraphics[width=0.49\textwidth]{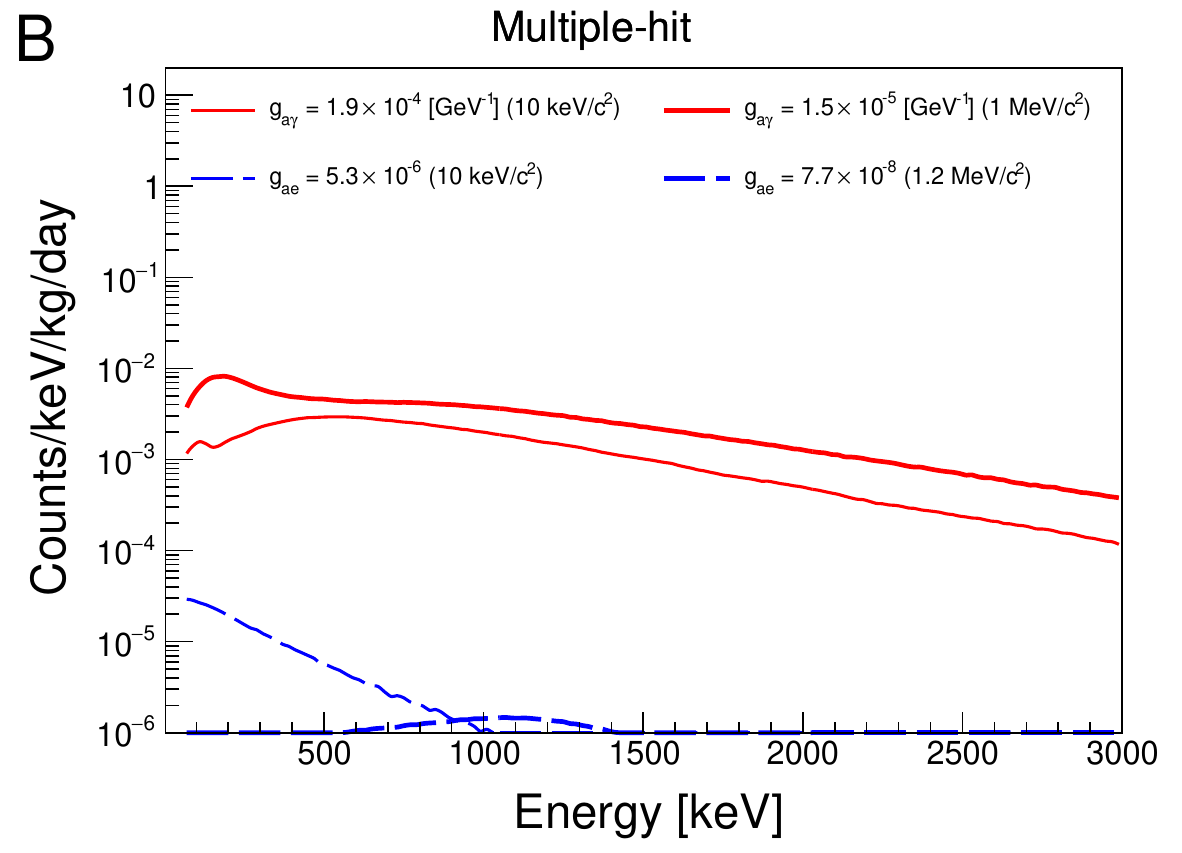} 
  \caption{Expected ALP signals in the NEON detector.
  The expected event rates from  ALP interactions in the NEON detector, incorporating detector responses, are presented for single-hit events (A) and multiple-hit events (B). Selected values of $g_{a\gamma}$ and $g_{ae}$ are shown for ALP masses of 10\,keV/c$^2$, 1\,MeV/c$^2$, and 1.2\,MeV/c$^2$. 
  }
  \label{figure:alpsim}
\end{figure}

ALP signals for each detection process in the mass range from 1\,eV/c$^{2}$ to 10\,MeV/c$^{2}$ were simulated. 
Calculated ALP fluxes in the reactor were generated isotropically within the reactor core volume, utilizing the geometry of the NEON detector to account for the direction and momentum of ALP events. Upon reaching the detector volume, which includes the liquid scintillator and NaI(Tl) crystals, ALP interactions were simulated, considering both scattering and decay processes for detector responses using Geant4-based simulations.   Figure~\ref{figure:alpsim} illustrates the expected ALP signals for a few benchmark scenarios in single-hit (A) and the multiple-hit (B) events. 

Typically, the $a \rightarrow e^{+}e^{-}$ and axio-electric processes produce electrons and positrons in the detector with energies below a few MeV. In such cases, most of the energy is absorbed by a single detector, resulting in single-hit events. Other processes, such as those involving MeV-energy photons, can deposit energy across multiple detectors through Compton scattering, leading to  multiple-hit events, as shown in Fig.~\ref{figure:alpsim}. Consequently, both single-hit and multiple-hit events are included in the ALP search analysis.

\end{document}